\documentclass[reprint, superscriptaddress, amsmath, amssymb, prl, floatfix]{revtex4-2}
\usepackage{graphicx}
\usepackage{bm}
\usepackage{amsmath}
\usepackage{xcolor}
\usepackage[colorlinks,linkcolor = blue, citecolor = blue,urlcolor = blue]{hyperref}
\usepackage{subcaption} 
\usepackage{physics}
\usepackage{amsmath}
\usepackage{amssymb}
\usepackage[font = small]{caption}
\usepackage{dsfont}

\makeatletter
\long\def\@makecaption#1#2{%
  \par
  \vskip\abovecaptionskip
  \begingroup
    \small\rmfamily
    \flushing
    \@make@capt@title{#1}{#2}\par
  \endgroup
  \vskip\belowcaptionskip}
\makeatother

\begin{document}

\title{Quantum Geometry in the Continuum: Solitons in Shallow Lattices}
\author{Koorosh Sadri}
\email{ksadri@psu.edu}
\affiliation{Department of Physics, The Pennsylvania State University, University Park, Pennsylvania 16802, USA}
\author{Mikael C. Rechtsman}
\email{mcrworld@psu.edu}
\affiliation{Department of Physics, The Pennsylvania State University, University Park, Pennsylvania 16802, USA}
\date{\today}

\begin{abstract}
    The quantum geometry of electronic, photonic, and atomic lattice systems quantifies the distance in Hilbert space between Bloch states at neighboring lattice momenta.  This quantity has profound implications for flat-band systems especially, characterizing surprising behavior such as superfluidity and superconductivity when the group velocity is zero and no transport would be expected for non-interacting particles.  However, when the band is not flat, the effects of quantum geometry are often intertwined with and partly masked by the band dispersion. Here, we show that in weakly interacting bosonic systems in the critical dimension (i.e., two dimensions for Kerr nonlinearity), the deviation from critical behavior due to the presence of the lattice is governed by the quantum geometry, which is directly proportional to the fourth-order dispersion. Furthermore, we identify the family of continuous lattice potentials that saturates the bound on the quantum metric for a given effective mass tensor. 
    
\end{abstract}

\maketitle

\emph{Introduction.} A common approach to describing waves propagating in lattices is to treat them as free particles with an effective mass.  A great deal of recent work related to Bloch state topology \cite{Haldane1987, KaneReviewOnTopological, TopologicalPhotonics} has shown how the dependence of Bloch states on lattice momentum - which is absent in the free-space description - leads to surprising new phenomena, such as topological insulators \cite{KaneMele2005,Molenkamp2006} the quantum anomalous Hall effect \cite{Haldane1987, Cui-Zu2013}, Weyl semimetals \cite{Vishwanath, Hasan2015, LuSoljacic2015, ReviewArmitage}, among many others.  Particularly surprising are the implications of the quantum geometry (or Fubini-Study metric) of the Bloch state manifold, which has a profound effect on flat-band systems, including superfluidity and superconductivity in the presence of flat bands; anomalous Landau level behavior, shift current, and orbital magnetism \cite{Yang20, TormaBernevig, torma15, ozawa}.

The nonlinear Schr\"odinger equation (NLSE), also called the Gross-Pitaevskii equation, provides a universal description of nonlinear wave dynamics in atomic gases, optical media, and effective continuum models of interacting bosons. In the presence of periodic potentials, and under attractive interactions, spatially localized bright stationary solitons bifurcate from Bloch band edges, and their shape, power, and stability are governed by the interplay between the band structure of the linear system and the nonlinear interaction \cite{Christodoulides1987, YaronSilberberg1998, Hulet02, Salomon02, FleischerNature2003}. The persistence of low-power solitons in periodic media and a systematic asymptotic framework connecting their properties to the underlying Bloch structure were established in Ref. \cite{weinstein}.  Particularly interesting is the critical dimension (i.e., the Townes soliton), because here power is independent of energy and critical collapse occurs \cite{MichaelWeinsteinStabilityPaper}.  

Gap solitons have been examined in a variety of settings, emphasizing the role of effective mass and higher-order dispersion in stability and bifurcation behavior \cite{aceves, ostrovskaya, pelinovsky, kevrekidisbook}. Paradigmatic experimental systems described by the NLSE, and in which many associated phenomena have been discovered, include ultracold atoms in optical lattices \cite{Greiner01, Morsch06, Ketterlevortices}, and photorefractive photonic lattices \cite{FleischerNature2003} and laser-written waveguide array systems \cite{Szameit2010}.  

Here, we study solitons bifurcating below the lowest Bloch band edge of the linear Schr\"odinger operator with a periodic potential. For weak periodic potentials in the critical dimension, we show that the leading correction to the band-edge slope relative to the free-space result is governed by the quantum metric, establishing a direct connection between quantum geometry and nonlinear localization. We further derive an identity relating the quantum metric to the fourth-order band dispersion in the shallow-lattice limit. For nonlinearities with critical exponent, this implies that the Vakhitov–Kolokolov stability signature \cite{vakhitov} is directly proportional to the quantum metric, implying that near the band edge, the power-energy derivative is positive and solitons near the band edge are therefore linearly unstable.  

Finally, we address the known inequality relating the effective mass tensor and the quantum metric \cite{iskin}. We identify a continuous family of periodic potentials that saturate this bound at the Brillouin-zone center. In the shallow-lattice limit, these correspond to sinusoidal potential with the same period as the lattice. In the opposite, deep-lattice limit, the potential evolves into arrays of nearly decoupled parabolic wells supporting localized harmonic-oscillator states.

\begin{figure}[t]
    \centering
    \begin{subfigure}[t]{0.48\columnwidth}
        \centering
        \includegraphics{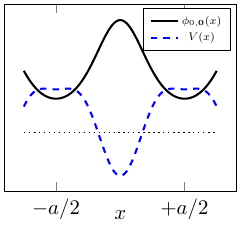}
        \caption{}
    \end{subfigure}\hfill
    \begin{subfigure}[t]{0.48\columnwidth}
        \centering
        \raisebox{6mm}{\includegraphics[height = 3.2cm]{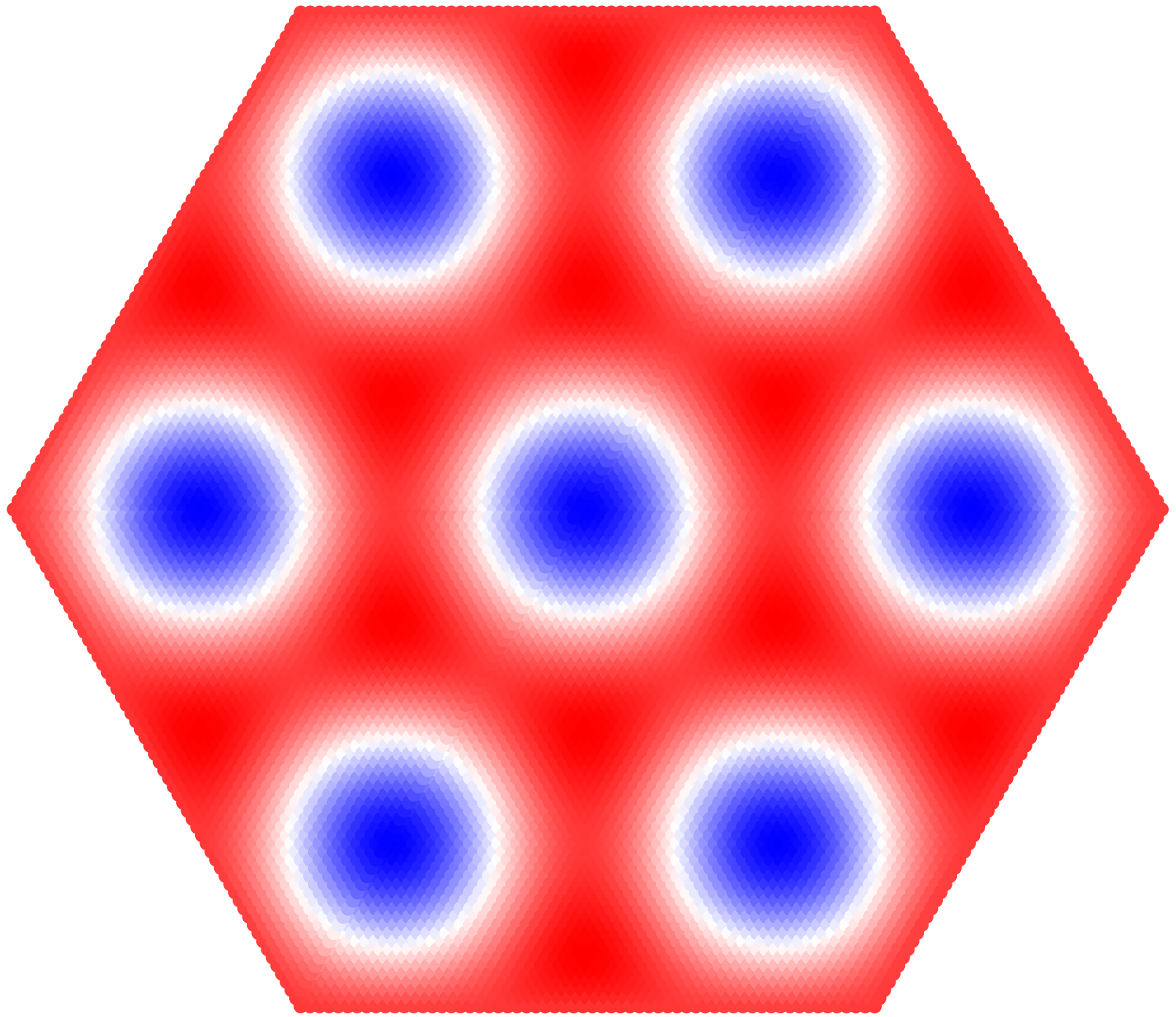}}
        \caption{}
    \end{subfigure}
    
    \hspace{-5mm}\begin{subfigure}[t]{1.1\columnwidth}
        \centering
        \includegraphics[scale = 0.7]{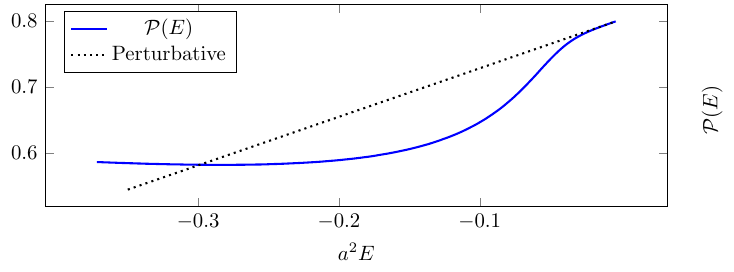}
        \caption{}
    \end{subfigure}
    
    \caption{(a) A sketch of the one dimensional ground state wavefunction $\phi_{0, 0}(x) = \exp[\Lambda_0\cos(2\pi x/a)]$ as well as the potential form setting $E_{0, \mathbf{0}} = 0$ for parameter $\Lambda_0 = 3/5$. (b) A heatmap of the two dimensional periodic potential corresponding to the ground state $\phi_{0, \mathbf{0}}(x) = \exp[\Lambda_0(\cos \mathbf{k}_1.\mathbf{x} \, + \, \cos\mathbf{k}_2.\mathbf{x}\, + \, \cos\mathbf{k}_3.\mathbf{x} )]$ with $a\mathbf{k}_1 = (1, -1/\sqrt{3})^T$, $a\mathbf{k}_2 = (1, +1/\sqrt{3})^T$, $a\mathbf{k}_3 = (0, 2/\sqrt{3})^T$ and $E_{0, \mathbf{0}} = 0$. (c) The power-energy curve for the one dimensional system from panel (a) with critical Kerr nonlinearity ($\sigma =2,\,  d = 1$)}
    \label{fig:1}
\end{figure}

\begin{figure}[!tbp]
    \centering
    \begin{subfigure}[t]{\columnwidth}
        \centering
        \includegraphics[]{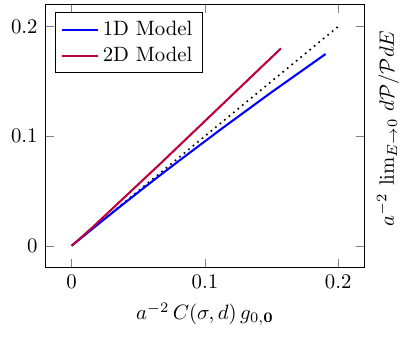}
        \caption{}
    \end{subfigure}\hfill
    \vspace{0.5em}
    \begin{subfigure}[t]{\columnwidth}
        \centering
        \hspace{3mm}
        \includegraphics[]{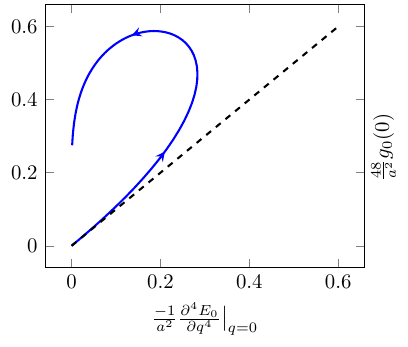}
        \caption{}
    \end{subfigure}
    
    \caption{(a) Power--energy slope $d\mathcal P/\mathcal{P}dE$ for the one- and two-dimensional isotropic potentials corresponding to $\phi_{0, 0}(x) = \exp[\Lambda_0\cos(2\pi x/a)]$ and $\phi_{0, \mathbf{0}}(x) = \exp[\Lambda_0(\cos \mathbf{k}_1.\mathbf{x} \, + \, \cos\mathbf{k}_2.\mathbf{x}\, + \, \cos\mathbf{k}_3.\mathbf{x} )]$ with $a\mathbf{k}_1 = (1, -1/\sqrt{3})^T$, $a\mathbf{k}_2 = (1, +1/\sqrt{3})^T$, $a\mathbf{k}_3 = (0, 2/\sqrt{3})^T$ shown in Fig.~1, plotted against the isotropic Fubini--Study quantum metric. Both cases have critical nonlinearity, $\sigma d = 2$. (b) Parametric plot of the quantum metric and the quartic dispersion as functions of the potential-strength parameter $\Lambda_0$ for the one dimensional system with $\phi_{0,0}(x) = \exp[\Lambda_0\cos(2\pi x/a)]$. The curve starts in the weak-potential limit, tangent to the linear relation predicted by Eq.~\eqref{metric-quartic}, and then continues toward stronger potentials where it approaches the $y$ axis}
    \label{fig:2}
\end{figure}

\emph{Quantum Geometric Corrections to the Townes Soliton}. We consider stationary soliton solutions of the nonlinear Schr\"odinger equation:
\begin{equation}\label{GPE}
\hat{H}\psi - E\psi - |\psi|^{2\sigma}\psi = 0,
\end{equation}
for the family of Hamiltonians $\hat H = -\tfrac12\nabla^2 + V(\mathbf{x})$, where $V(\mathbf{x})$ is periodic. This class of Hamiltonians may as well be specified through their eigenstates and spectra. Due to the discrete translational symmetry arising from the periodicity of $V(\mathbf{x})$, the spectrum consists of Bloch bands $E_\nu(\mathbf{q})$ with eigenstates $e^{i\mathbf{q}\cdot\mathbf{x}}\phi_{\nu,\mathbf{q}}(\mathbf{x})$, where $\phi_{\nu,\mathbf{q}}$ are normalized periodic functions defined over a unit cell $D$ and $\nu$ is the discrete band number. The dispersion — i.e., the dependence of the band energy $E_\nu(\mathbf{q})$ on the quasi-momentum $\mathbf{q}$ — plays an important role in determining the soliton shape and power. Similarly, as we will see, the dependence of the eigenstates $\phi_{\nu, \mathbf{q}}$ on $\mathbf{q}$ will also lead to corrections to the shape and power of these nonlinear states. This latter $\mathbf{q}$-dependence is best quantified through the gauge-invariant quantum geometric tensor
\begin{equation}
    (Q_{\nu})_{ij} = \bra{\pdv{\phi_{\nu, \mathbf{q}}}{q_i}} \Big(\mathds{1} - \ket{\phi_{\nu, \mathbf{q}}}\bra{\phi_{\nu, \mathbf{q}}}\Big)\ket{\pdv{\phi_{\nu, \mathbf{q}}}{q_j}}.
\end{equation}
The quantum geometric tensor is usually decomposed to its real and imaginary parts as $Q_\nu = g_\nu - \frac{i}{2}\Omega_\nu$. The imaginary part is the 2-form known as the Berry curvature. The real part, $g_\nu$, is a proper, positive definite Riemannian metric tensor that coincides with the Fubini-Study distance of adjacent eigenstates.

In the free-particle case $V\equiv0$, soliton solutions exist \cite{Townes, MichaelWeinsteinStabilityPaper} for $E < 0$. Defining, the power $\mathcal{P} = \int |\psi|^2\,d\mathbf{x}$ to characterize the norm of the nonlinear mode, these free solitons satisfy the scaling law $\mathcal P_\mathrm{Free}(E)\propto(-E)^{1/\sigma-d/2}$.  The dimension $d = 2/\sigma$ is called critical, because the energy does not depend on power, and the system is scale-invariant, leading to critical collapse~\cite{MichaelWeinsteinStabilityPaper}.

 In the presence of a non-zero potential, the energy $E$ of the soliton states fall inside spectral gaps of the linear Hamiltonian and bifurcate from band edges. We choose the additive constant of the potential so that the bottom band edge satisfies $E_0(\mathbf{0}) = 0$. Representative periodic potentials and a corresponding power–energy behavior of solitons bifurcating from their ground states are illustrated in Fig. \ref{fig:1}.

We study nonlinear solutions to eq. \eqref{GPE} with energy $E = -h^2/2$, where $h$ measures the detuning from the bottom band edge. Building on the asymptotic framework of Ref.~\cite{weinstein}, we use the multiband envelope ansatz
\begin{equation}\label{expansion}
    \psi(\mathbf{x}; -h^2/2) = h^{\frac{1}{\sigma} - d}\sum_\nu\int_B \frac{d\mathbf{q}}{\sqrt{|B|}}\,e^{i\mathbf{q}\cdot\mathbf{x}}\phi_{\nu,\mathbf{q}}(\mathbf{x})\,\tilde F_\nu(\mathbf{q}/h, h).
\end{equation}
Here $B$ is the first Brillouin zone and $\tilde{F}$ are Fourier transforms of spatial envelope functions $F$. Once we find the envelope functions, the soliton power at the energy $E = -h^2/2$ is found as $\mathcal{P}(E = -h^2/2) = h^{\frac{2}{\sigma} - d}\sum_\nu\int d\mathbf{k}\,|\tilde F_\nu(\mathbf{k};h)|^2$ where the $\mathbf{k}$ integration domain is strictly $B/h$, but for smooth spatial envelopes $F_\nu(\mathbf{X},h)$ extending the integral to $\mathbb{R}^d$ is superpolynomially accurate in $h$.

We solve equation \eqref{GPE} for the envelope functions order by order in $h$ by expanding the real-space envelopes
$F_\nu(\mathbf{X},h) = \sum_{n=0}^\infty h^n F_{\nu,n}(\mathbf{X})$.
At zeroth order, all higher bands with $\nu > 0$ have vanishing amplitudes, and for $\nu=0$
the envelope depends on the band structure only through the effective mass tensor
of the lowest band,
\begin{equation}
    F_{0,0}(\mathbf{X}) = \frac{\sqrt{|B|}}{(2I_0)^{1/2\sigma}(2\pi)^{d/2}}f_\sigma(\sqrt{m}\,\mathbf{X}).
\end{equation}
Here $\sqrt{m}$ denotes the matrix root of the mass tensor defined by the quadratic expansion
$E_0(\mathbf{q})=\frac12 m^{-1}_{ij}q_iq_j+\cdots$.
The function $f_\sigma$ is the unique positive, radially symmetric, and decaying soliton satisfying
$\nabla^2 f_\sigma - f_\sigma + f_\sigma^{2\sigma+1} = 0$,
and $I_0=\int_D |\phi_{0,\mathbf{0}}(\mathbf{x})|^{2\sigma+2}\,d\mathbf{x}$. This result is consistent with Ref.~\cite{weinstein} for solitons bifurcating below the lowest band edge of periodic Schr\"odinger operators. More generally, the same leading-order structure applies near any band edge in any gap, including discrete systems, where the unit-cell integral is replaced by a sum over internal lattice indices, $I_{0,\mathrm{discrete}}\rightarrow \sum_i |\phi_{0,i}(\mathbf{0})|^{2(\sigma+1)}$.

Higher order corrections arise from two distinct sources: (i) deviations of the lowest-band dispersion from the quadratic form $E_0(\mathbf{q})\simeq q^2/2m$, and (ii) quantum geometry, encoded in the $\mathbf{q}$ dependence of the Bloch states $\phi_{\nu,\mathbf{q}}(\mathbf{x})$. At order $\mathcal{O}(h)$, all corrections vanish due to the time reversal symmetry of the linear Hamiltonian. As a result, the leading nontrivial contributions arise at order $\mathcal{O}(h^2)$, where quantum geometric effects enter. At $\mathcal{O}(h^2)$ corrections arise from two contributions: the quartic dispersion tensor $(\partial^4 E_0/\partial \mathbf{q}^4)|_{\mathbf{q}=\mathbf{0}}$ and geometric mixing terms involving second derivatives of the Bloch functions, namely the two integrals $\int_D \phi_{0,\mathbf{q}}^*\,\partial_{\mathbf{q}}^2\phi_{0,\mathbf{q}}\,d\mathbf{x}$ and $\int_D |\phi_{0,\mathbf{q}}|^{2\sigma}\phi_{0,\mathbf{q}}^*\,\partial_{\mathbf{q}}^2\phi_{0,\mathbf{q}}\,d\mathbf{x}$, both evaluated at $\mathbf{q} = \mathbf{0}$.

At this point, we focus on the weak-potential limit, characterized by small logarithmic fluctuations of the lowest-band Bloch state, $\Lambda \equiv \log\phi_{0,\mathbf{0}} - \langle\log\phi_{0,\mathbf{0}}\rangle$; where the average $\langle.\rangle$ denotes spatial averaging over a unit cell. Formulating the relevant mixing integrals, and the quartic dispersion tensor in terms of the logarithmic fluctuation $\Lambda(\mathbf{x})$ we find that all of the $\mathcal{O}(h^2)$ contributions are proportional to the quantum metric $g_{0,ij}(\mathbf{0}) \approx -4\langle\Lambda\,\Delta^{-2}\partial_i\partial_j\Lambda\rangle$. Notably, the quartic dispersion tensor admits the representation $\partial^4E_0/\partial q_i\partial q_j\partial q_k\partial q_l|_{\mathbf{q}=\mathbf{0}} \approx 192\langle\Lambda\,\Delta^{-3}\partial_i\partial_j\partial_k\partial_l\Lambda\rangle$, from which one obtains the shallow potential identity
\begin{equation}\label{metric-quartic}
    g_{0, ij}(\mathbf{0}) = -\frac{1}{48}\, \sum_k\left(\frac{\partial^4E_0}{\partial q_i\partial q_j \partial q_k \partial q_k}\right)_{\mathbf{q} = \mathbf{0}} + \mathcal{O}(\Lambda^3).
\end{equation}
Details of the derivation of eq. \eqref{metric-quartic} as well as the corrections to the shape and power of the solitons, are provided in the Supplemental Material, Section 1. The relation between the quartic dispersion tensor and the quantum metric is verified numerically in Fig.~2.

Here, we quote the resulting correction to the power-energy slope in the isotropic case, where both the quantum metric and the quartic dispersion tensor reduce to scalars parametrized by $g_{0, \mathbf{0}}$ as $ds^2 = g_ {0, \mathbf{0}} |d\mathbf{q}|^2$, and $E_0(\mathbf{q}) = |\mathbf{q}|^2/2m - 6g_{0, \mathbf{0}}|\mathbf{q}|^4/(d + 2) + \cdots$. In this limit we find that the leading order correction to the power-energy slope depends only on $g_{0, \mathbf{0}}$. 
\begin{equation}\label{gprime}
    \frac{d\mathcal{P}}{\mathcal{P}dE} = \frac{1/\sigma - d/2}{E} + C(\sigma, d)\,g_{0, \mathbf{0}} + \mathcal O(E,\Lambda^3),
\end{equation}
where the first term reproduces the familiar, potential-free scaling, $g_{0, \mathbf{0}}$ denotes the Fubini--Study metric evaluated at the band minimum, and $C(\sigma, d)$ is a numerical coefficient. The resulting linear dependence of $d\mathcal P/\mathcal{P}dE$ on the quantum metric is confirmed numerically in Fig.~2.

In the two critical cases $(\sigma, d) = (1, 2)$ and $(2, 1)$, we find that the dimension-dependent coefficients satisfy $C(1, 2)\approx 102.09 > 0$ and $C(2, 1) \approx 133.6 > 0$, implying a positive band-edge slope $d\mathcal P/dE$ in the weak-potential, isotropic regime. According to the Vakhitov--Kolokolov criterion, this corresponds to spectral instability of the associated band-edge solitons. This is in accordance with Weinstein’s analysis~\cite{weinstein} of critical band-edge bifurcations, where the perturbative correction to Eq.~6 was conjectured to be positive, and provides an explicit realization of that conjecture in terms of the Fubini--Study quantum metric for shallow isotropic lattices.  The main result of this section is the direct proportionality between the quantum metric, the fourth-order dispersion, and the slope of the power-energy curve for the case of shallow potentials.  In the critical case, this provides a new mathematical intuition for the deviation from critical Townes behavior, and therefore soliton instability near the band edge.

\begin{figure} [!tbp]
    \centering
    \hspace{-4mm}
    \begin{subfigure}{0.9\linewidth}
        \includegraphics[]{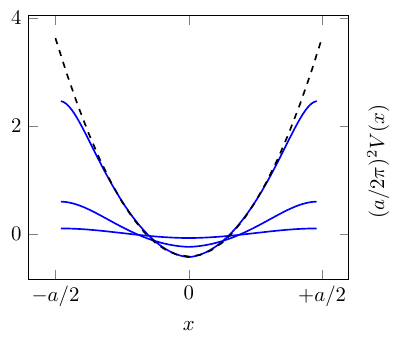}
        \caption{}
    \end{subfigure}
    \vspace{0.5em}
    \begin{subfigure}{0.9\linewidth}
        \includegraphics[]{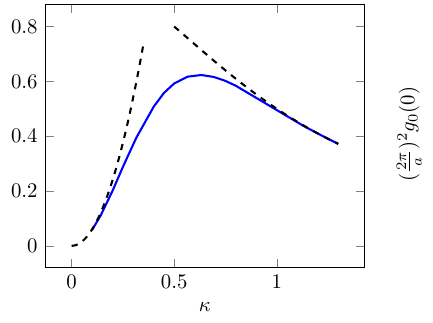}
        \caption{}
    \end{subfigure}
    
    \caption{(a) The shape of the potentials saturating the mass - metric bound in eq.\eqref{mass-metric} for 3 different values of $\kappa$: $0.1, \,0.4, \,0.85$ (larger $\kappa$ corresponds to stronger potential). The dashed curve consists of two quadratic wells. (b) Quantum metric of the lowest band at $q = 0$ for the saturating potential family, shown as a function of the potential-strength parameter $\kappa$. The drawn asymptotes are $g_0(0)\approx 6(\kappa a/2\pi)^2$ for the shallow limit and $g_0(0) \approx (a/2\pi)^2/(1 + \kappa^2)$}
    \label{fig:3}
\end{figure}

\emph{Quantum Geometry and the Effective Mass Tensor.} We have seen above that the quantum metric can be used in the case of critical Townes nonlinearity to diagnose the deviations from critical behavior.  Below, we examine the upper bound on the quantum geometry for continuous periodic systems with scalar potentials, in terms of the effective mass.  This therefore provides a bound on the slope of the power-energy curve (see Eq.~(5)) for such systems in terms of their effective mass tensor.  Furthermore, we find a one-parameter family of potentials that saturates this bound.

A discrete lattice model with arbitrary hopping amplitudes possesses sufficient freedom to allow independent tuning of its quantum geometry and energy band structure. By contrast, continuous periodic systems described by Schr\"odinger-type Hamiltonians of the form $H = -\frac12\nabla^2 + V(\mathbf{x})$ are more constrained: it is impossible to independently tune the dispersion and quantum geometry. The group velocity, effective mass, and quantum geometric tensor can all be expressed in terms of the Hermitian, vector valued velocity matrix
\begin{equation}
    \boldsymbol{\beta}_{\mu\nu}(\mathbf{q}) = \bra{\phi_{\mu,\mathbf{q}}}-i\boldsymbol{\nabla} + \mathbf{q}\ket{\phi_{\nu,\mathbf{q}}}.
\end{equation}
The group velocity of a band $\mu$ at any point in the Brillouin zone is then $\mathbf{v}_\mu(\mathbf{q}) \equiv \partial E_\mu/\partial\mathbf{q}= \boldsymbol{\beta}_{\mu\mu}(\mathbf{q})$. The mass tensor of a band $\mu$ is
\begin{equation}\label{mass}
    (m^{-1}_\mu)_{ij} = \pdv{^2E_\mu}{q_i \partial q_j} = \delta_{ij} - \sum_{\nu\neq \mu} \frac{\beta_{\mu\nu, i}\beta_{\mu\nu,j}^* + \beta_{\mu\nu,i}^*\beta_{\mu\nu,j}}{E_\nu - E_\mu}
\end{equation}
and the quantum geometric tensor for the $\mu$th band is
\begin{equation}
    (Q_\mu)_{ij} = (g_\mu)_{ij} - \frac{i}{2}(\Omega_\mu)_{ij} = \sum_{\nu\neq \mu}\frac{\beta_{\mu\nu,i}\beta_{\mu\nu,j}^*}{(E_\mu - E_\nu)^2}
\end{equation}
where $g_\mu$ is the quantum metric tensor and $\Omega_\mu$, the Berry curvature. 

A corollary of this connection between the mass tensor and the quantum metric is that, for the lowest band, the effective mass tensor is always larger than or equal to the free mass (unity in our units). This is because the subtracted term in eq. \eqref{mass} is positive definite when all the denominators are positive. Furthermore, the quantum metric is bounded from above by the difference between the free and effective mass tensors through the matrix inequality \cite{iskin},
\begin{equation}\label{mass-metric}
    g_0(\mathbf q) \leq \frac{\mathds{1} - m_0^{-1}(\mathbf{q})}{2\big(E_1(\mathbf q)-E_0(\mathbf q)\big)}.
\end{equation}

This bound is saturated if and only if all matrix elements $\boldsymbol{\beta}_{\mu0}$ vanish for $\mu > 1$. Equivalently,
\begin{equation}
    (-i\boldsymbol{\nabla} + \mathbf{q})\ket{\phi_{0,\mathbf q}} = \boldsymbol{\beta}_{00}\ket{\phi_{0,\mathbf q}} + \boldsymbol{\beta}_{10}\ket{\phi_{1,\mathbf q}} .
\end{equation}
We determine the class of periodic potentials that satisfy this condition at $\mathbf q = \mathbf 0$ while setting $E_0(\mathbf 0) = 0$. As shown in the Supplemental Material, Section 2, this implies that saturating the bound given by Eq. (10) leads to the nonlinear eigenvalue equation for the logarithmic fluctuation $\Lambda(\mathbf{x}) = \log(\phi_{0, \mathbf{0}}(\mathbf{x})) - \langle \log(\phi_{0, \mathbf{0}}(\mathbf{x}))\rangle$ 
\begin{equation}\label{saturator}
    [\nabla^2 + 2E_1(\mathbf{0})]\Lambda = \langle|\boldsymbol{\nabla}\Lambda|^2\rangle - |\boldsymbol{\nabla}\Lambda|^2 .
\end{equation}
Recall that $\Lambda$ quantifies the deviation of the Bloch mode from a pure plane wave, which has zero quantum geometry. The one-parameter family of saturating solutions is controlled by a dimensionless parameter $\kappa$, which sets the strength of the periodic potential by fixing the band energy gap: $E_1(\mathbf{0}) = (1 + \kappa^2)E_1^{\mathrm{free}}(\mathbf{0})$. Therefore, $\kappa \ll 1$ corresponds to the weak-potential regime, while $\kappa \gg 1$ corresponds to the deep-lattice regime. For the one dimensional system, and in the shallow-lattice limit $\kappa \ll 1$, the potential takes the form
\begin{equation}
    V(x) \approx -\frac{1}{2}\left(\frac{2\pi}{a}\right)^2 \sqrt{3}\,\kappa \cos\!\left(\frac{2\pi x}{a}\right),
\end{equation}
and the effective mass and quantum metric behave as $m_0(0) \approx 1 + 6\kappa^2$, and $g_0(0) \approx 6 (\kappa a/2\pi)^2$. As $\kappa$ grows, the saturating potential turns into an array of quadratic wells hosting localized, decoupled Gaussian modes. This leads to exponentially increasing effective mass, and hence small dispersion, while the quantum metric decays as $g_0(0) \propto \frac{1}{1+\kappa^2}$. Plots of three instances of these potentials for different $\kappa$ values are drawn in Fig.~3. Furthermore, the dependence of the optimal $g_0(0)$ on $\kappa$ is exhibited.

The same inequality \eqref{mass-metric} generalizes to Hamiltonians with a periodic magnetic vector potential $H = -\frac12[\boldsymbol{\nabla} - i\mathbf{A}(\mathbf{x})]^2 + V(\mathbf{x})$. The conditions required for saturation are reported in the Supplemental Material, Section 2. As a simple example, one may choose the two dimensional system with $H = -\frac12\nabla^2 + iA_0\cos(2\pi x/a)\partial_y$, which yields a nonzero magnetic field and saturates the bound. This solution depends on only one of the two spatial coordinates.

\emph{Conclusion.} In conclusion, we demonstrated that quantum geometry plays a direct role in the formation and stability of band-edge solitons in periodic systems governed by the nonlinear Schr\"odinger equation, even in the dispersive (non-flat-band) case. Using a multiband perturbative expansion, we showed that in the weak-potential regime the leading correction to the power–energy relation is controlled by the Fubini--Study quantum metric, and we derived an explicit identity connecting the metric to the quartic band dispersion in the shallow-lattice limit. When the dispersion is isotropic at the band edge, this yields a compact expression for $d\mathcal P/(\mathcal P\,dE)$, and in the critical cases $(\sigma,d)=(1,2)$ and $(2,1)$ fixes the Vakhitov--Kolokolov criterion, implying spectral instability of band-edge solitons. We further identified a continuous family of periodic potentials that saturates the general mass–metric inequality, interpolating between a cosine potential in the weak potential limit and deep arrays of parabolic wells. These results establish quantum geometry as a fundamental control parameter for nonlinear localization and provide a concrete framework for relating soliton properties to geometric features of Bloch bands, relevant to optical lattices of ultracold atoms and photonic lattices.

\emph{Acknowledgements.} We thank Michael Weinstein for useful discussions.  We acknowledge the support of the Air Force Office of Scientific Research under the MURI program, grant number FA9550-22-1-0339.

\bibliography{references}

@article{iskin,
    author       = {Menderes Iskin},
    title        = {Quantum-metric contribution to the pair mass in spin-orbit coupled Fermi superfluids},
    journal      = {Phys. Rev. A},
    volume       = {97},
    number       = {3},
    pages        = {033625},
    year         = {2018},
    doi          = {10.1103/PhysRevA.97.033625}
}

@article{vakhitov,
    title = {Stationary solutions of the wave equation in a medium with nonlinearity saturation}, 
    author = {Vakhitov, N. G. and Kolokolov, A. A.},
    journal = {Radiophys. Quantum Electron.},
    volume = {16},
    pages = {783},
    year = {1973}
}

@inproceedings{aceves,
author = {Alejandro B. Aceves and Stefan Wabnitz},
booktitle = {Annual Meeting Optical Society of America},
journal = {Annual Meeting Optical Society of America},
pages = {TUF1},
title = {Self-induced transparency solitons in nonlinear refractive periodic media},
year = {1989},
url = {https://opg.optica.org/abstract.cfm?URI=OAM-1989-TUF1},
}

@article{ostrovskaya,
  title = {Matter-Wave Gap Solitons in Atomic Band-Gap Structures},
  author = {Ostrovskaya, Elena A. and Kivshar, Yuri S.},
  journal = {Phys. Rev. Lett.},
  volume = {90},
  issue = {16},
  pages = {160407},
  numpages = {4},
  year = {2003},
  month = {Apr},
  publisher = {American Physical Society},
  doi = {10.1103/PhysRevLett.90.160407},
  url = {https://link.aps.org/doi/10.1103/PhysRevLett.90.160407}
}

@article{pelinovsky,
  title = {Bifurcations and stability of gap solitons in periodic potentials},
  author = {Pelinovsky, Dmitry E. and Sukhorukov, Andrey A. and Kivshar, Yuri S.},
  journal = {Phys. Rev. E},
  volume = {70},
  issue = {3},
  pages = {036618},
  numpages = {17},
  year = {2004},
  month = {Sep},
  publisher = {American Physical Society},
  doi = {10.1103/PhysRevE.70.036618},
  url = {https://link.aps.org/doi/10.1103/PhysRevE.70.036618}
}

@article{weinstein,
    author = {Ilan, B. and Weinstein, M. I.},
    title = {Band-Edge Solitons, Nonlinear Schrödinger/Gross–Pitaevskii Equations, and Effective Media},
    journal = {Multiscale Modeling \& Simulation},
    volume = {8},
    number = {4},
    pages = {1055-1101},
    year = {2010},
    URL = {https://doi.org/10.1137/090769417}
}

@article{torma15,
  author       = {Peotta, Sebastiano and T\"{o}rm\"{a}, P\"{a}ivi},
  title        = {Superfluidity in topologically nontrivial flat bands},
  journal      = {Nature Communications},
  volume       = {6},
  pages        = {8944},
  year         = {2015},
  doi          = {10.1038/ncomms9944},
  url          = {https://www.nature.com/articles/ncomms9944}
}

@article{ozawa,
  title = {Extracting the quantum metric tensor through periodic driving},
  author = {Ozawa, Tomoki and Goldman, Nathan},
  journal = {Phys. Rev. B},
  volume = {97},
  issue = {20},
  pages = {201117},
  numpages = {6},
  year = {2018},
  month = {May},
  publisher = {American Physical Society},
  doi = {10.1103/PhysRevB.97.201117},
  url = {https://link.aps.org/doi/10.1103/PhysRevB.97.201117}
}

@article{Haldane1987,
  title = {Model for a Quantum Hall Effect without Landau Levels: Condensed-Matter Realization of the "Parity Anomaly"},
  author = {Haldane, F. D. M.},
  journal = {Phys. Rev. Lett.},
  volume = {61},
  issue = {18},
  pages = {2015--2018},
  numpages = {0},
  year = {1988},
  month = {Oct},
  publisher = {American Physical Society},
  doi = {10.1103/PhysRevLett.61.2015},
  url = {https://link.aps.org/doi/10.1103/PhysRevLett.61.2015}
}

@article{KaneReviewOnTopological,
  title = {Colloquium: Topological insulators},
  author = {Hasan, M. Z. and Kane, C. L.},
  journal = {Rev. Mod. Phys.},
  volume = {82},
  issue = {4},
  pages = {3045--3067},
  numpages = {0},
  year = {2010},
  month = {Nov},
  publisher = {American Physical Society},
  doi = {10.1103/RevModPhys.82.3045},
  url = {https://link.aps.org/doi/10.1103/RevModPhys.82.3045}
}

@article{TopologicalPhotonics,
  title = {Topological photonics},
  author = {Ozawa, Tomoki and Price, Hannah M. and Amo, Alberto and Goldman, Nathan and Hafezi, Mohammad and Lu, Ling and Rechtsman, Mikael C. and Schuster, David and Simon, Jonathan and Zilberberg, Oded and Carusotto, Iacopo},
  journal = {Rev. Mod. Phys.},
  volume = {91},
  issue = {1},
  pages = {015006},
  numpages = {76},
  year = {2019},
  month = {Mar},
  publisher = {American Physical Society},
  doi = {10.1103/RevModPhys.91.015006},
  url = {https://link.aps.org/doi/10.1103/RevModPhys.91.015006}
}

@article{KaneMele2005,
  title = {Quantum Spin Hall Effect in Graphene},
  author = {Kane, C. L. and Mele, E. J.},
  journal = {Phys. Rev. Lett.},
  volume = {95},
  issue = {22},
  pages = {226801},
  numpages = {4},
  year = {2005},
  month = {Nov},
  publisher = {American Physical Society},
  doi = {10.1103/PhysRevLett.95.226801},
  url = {https://link.aps.org/doi/10.1103/PhysRevLett.95.226801}
}

@article{Molenkamp2006,
    author = {Markus Konig and Steffen Wiedmann and Christoph Brune and Andreas Roth and Hartmut Buhmann and Laurens W Molenkamp and Xiao-Liang Qi and Shou-Cheng Zhang},
    title = {Quantum spin Hall insulator state in HgTe quantum wells},
    journal = {Science},
    year = {2007},
    url = {https://www.science.org/doi/abs/10.1126/science.1148047}
}

@article{Cui-Zu2013,
    author = {Cui-Zu Chang and Jinsong Zhang and Xiao Feng and Jie Shen and Zuocheng Zhang and Minghua Guo and Kang Li and Yunbo Ou and Pang Wei and Li-Li Wang and Zhong-Qing Ji and Yang Feng and Shuaihua Ji and Xi Chen and Jinfeng Jia and Xi Dai and Zhong Fang and Shou-Cheng Zhang and Ke He and Yayu Wang and Li Lu and Xu-Cun Ma and Qi-Kun Xue},
    title = {Experimental observation of the quantum anomalous Hall effect in a magnetic topological insulator},
    journal = {Science},
    year = {2013},
    url = {https://www.science.org/doi/abs/10.1126/science.1234414}
}

@article{Vishwanath,
  title = {Topological semimetal and Fermi-arc surface states in the electronic structure of pyrochlore iridates},
  author = {Wan, Xiangang and Turner, Ari M. and Vishwanath, Ashvin and Savrasov, Sergey Y.},
  journal = {Phys. Rev. B},
  volume = {83},
  issue = {20},
  pages = {205101},
  numpages = {9},
  year = {2011},
  month = {May},
  publisher = {American Physical Society},
  doi = {10.1103/PhysRevB.83.205101},
  url = {https://link.aps.org/doi/10.1103/PhysRevB.83.205101}
}

@article{Hasan2015,
    author = {Su-Yang Xu and Ilya Belopolski and Nasser Alidoust and Madhab Neupane and Chenglong Zhang and Raman Sankar and Shin-Ming Huang and Chi-Cheng Lee and Guoqing Chang and BaoKai Wang and Guang Bian and Hao Zheng and Daniel S. Sanchez and Fangcheng Chou and Hsin Lin and Shuang Jia and M. Zahid Hasan},
    title = {Discovery of a Weyl fermion semimetal and topological Fermi arcs},
    journal = {Science},
    year = {2015},
    url = {https://www.science.org/doi/10.1126/science.aaa9297}
}

@article{LuSoljacic2015,
    author = {Lu, Ling and Joannopoulos, John D. and Soljacic, Marin},
    title = {Topological photonics},
    journal = {Nature Photonics},
    year = {2014},
    url = {https://dspace.mit.edu/handle/1721.1/98067}
}

@article{ReviewArmitage,
  title = {Weyl and Dirac semimetals in three-dimensional solids},
  author = {Armitage, N. P. and Mele, E. J. and Vishwanath, Ashvin},
  journal = {Rev. Mod. Phys.},
  volume = {90},
  issue = {1},
  pages = {015001},
  numpages = {57},
  year = {2018},
  month = {Jan},
  publisher = {American Physical Society},
  doi = {10.1103/RevModPhys.90.015001},
  url = {https://link.aps.org/doi/10.1103/RevModPhys.90.015001}
}

@article{Christodoulides1987,
    author = {D. N. Christodoulides and R. I. Joseph},
    journal = {Opt. Lett.},
    number = {9},
    pages = {794--796},
    title = {Discrete self-focusing in nonlinear arrays of coupled waveguides},
    volume = {13},
    month = {Sep},
    year = {1988},
    url = {https://opg.optica.org/ol/abstract.cfm?URI=ol-13-9-794},
    doi = {10.1364/OL.13.000794}
}

@article{YaronSilberberg1998,
  title = {Discrete Spatial Optical Solitons in Waveguide Arrays},
  author = {Eisenberg, H. S. and Silberberg, Y. and Morandotti, R. and Boyd, A. R. and Aitchison, J. S.},
  journal = {Phys. Rev. Lett.},
  volume = {81},
  issue = {16},
  pages = {3383--3386},
  numpages = {0},
  year = {1998},
  month = {Oct},
  publisher = {American Physical Society},
  doi = {10.1103/PhysRevLett.81.3383},
  url = {https://link.aps.org/doi/10.1103/PhysRevLett.81.3383}
}

@article{Hulet02,
    author = {Kevin E. Strecker and Guthrie B. Partridge and Andrew G. Truscott and Randall G. Hulet},
    title = {Formation and propagation of matter-wave soliton trains},
    journal = {Nature},
    year = {2002},
    url = {https://www.nature.com/articles/nature747}
}

@article{Salomon02,
    author = {L Khaykovich and F Schreck and Gabriele Ferrari and Thomas Bourdel and Julien Cubizolles and Lincoln D Carr and Yvan Castin and Christophe Salomon},
    title = {Formation of a matter-wave bright soliton},
    journal = {Science},
    year = {2002},
    url = {https://www.science.org/doi/abs/10.1126/science.1071021}
}

@article{FleischerNature2003,
    author = {Jason W. Fleischer and Mordechai Segev and Nikolaos K. Efremidis and Demetrios N. Christodoulides },
    title = {Observation of two-dimensional discrete solitons in optically induced nonlinear photonic lattices},
    journal = {Nature},
    year = {2003},
    url = {https://www.nature.com/articles/nature01452}
}

@book{kevrekidisbook,
    author = {Panayotis G. Kevrekidis},
    title = {The Discrete Nonlinear Schrödinger Equation},
    publisher = {Springer},
    year = {2009}
}

@article{Ketterlevortices,
    author = {Jamil R Abo-Shaeer and Chandra Raman and Johnny M Vogels and Wolfgang Ketterle},
    title = {Observation of Vortex Lattices in Bose-Einstein Condensates},
    journal = {Science},
    year = {2001}
}

@article{Szameit2010,
    doi = {10.1088/0953-4075/43/16/163001},
    url = {https://doi.org/10.1088/0953-4075/43/16/163001},
    year = {2010},
    month = {jul},
    publisher = {},
    volume = {43},
    number = {16},
    pages = {163001},
    author = {Szameit, Alexander and Nolte, Stefan},
    title = {Discrete optics in femtosecond-laser-written photonic structures},
    journal = {Journal of Physics B: Atomic, Molecular and Optical Physics}
}

@article{Townes,
  title = {Self-Trapping of Optical Beams},
  author = {Chiao, R. Y. and Garmire, E. and Townes, C. H.},
  journal = {Phys. Rev. Lett.},
  volume = {13},
  issue = {15},
  pages = {479--482},
  numpages = {0},
  year = {1964},
  month = {Oct},
  publisher = {American Physical Society},
  doi = {10.1103/PhysRevLett.13.479},
  url = {https://link.aps.org/doi/10.1103/PhysRevLett.13.479}
}

@article{MichaelWeinsteinStabilityPaper,
    author = {Weinstein, Michael I.},
    title = {Lyapunov stability of ground states of nonlinear dispersive evolution equations},
    journal = {Communications on Pure and Applied Mathematics},
    volume = {39},
    number = {1},
    pages = {51-67},
    url = {https://onlinelibrary.wiley.com/doi/abs/10.1002/cpa.3160390103},
    year = {1986}
}

@article{Morsch06,
  title = {Dynamics of Bose-Einstein condensates in optical lattices},
  author = {Morsch, Oliver and Oberthaler, Markus},
  journal = {Rev. Mod. Phys.},
  volume = {78},
  issue = {1},
  pages = {179--215},
  numpages = {0},
  year = {2006},
  month = {Feb},
  publisher = {American Physical Society},
  doi = {10.1103/RevModPhys.78.179},
  url = {https://link.aps.org/doi/10.1103/RevModPhys.78.179}
}

@article{Greiner01,
    author = {Markus Greiner and Olaf Mandel and Tilman Esslinger and Theodor W. Hänsch and Immanuel Bloch},
    title = {Quantum phase transition from a superfluid to a Mott insulator in a gas of ultracold atoms},
    journal = {Nature},
    year = 2001
}

@article{Yang20,
    author = {Jun-Won Rhim and Kyoo Kim and Bohm-Jung Yang},
    title = {Quantum distance and anomalous Landau levels of flat bands},
    journal = Nature,
    year = 2020
}

@article{TormaBernevig,
    author = {P. Törmä and S. Peotta and B. A. Bernevig},
    title = {Superfluidity and Quantum Geometry in Twisted Multilayer Systems},
    journal = {Nature Reviews Physics},
    year = 2022
}
\bibliographystyle{apsrev4-2.bst}

\clearpage
\onecolumngrid
\clearpage

\begin{center}
    \textbf{\large Supplemental Material}
\end{center}

\vspace{1em}

\setcounter{section}{0}
\setcounter{equation}{0}
\setcounter{figure}{0}
\setcounter{table}{0}

\renewcommand{\thesection}{S\arabic{section}}
\renewcommand{\theequation}{S\arabic{equation}}
\renewcommand{\thefigure}{S\arabic{figure}}
\renewcommand{\thetable}{S\arabic{table}}

\section{Section 1: Formulation in Terms of Logarithmic Fluctuations}

The most natural way to parametrize the periodic Hamiltonian system
\begin{equation}\label{hamiltonian}
    H = -\frac12\nabla^2 + V(\mathbf{x}) 
\end{equation}
is to specify the potential field $V(\mathbf{x})$. However, this makes it impossible to find the explicit form of the Bloch states or the energy band structure. Conversely, we can parametrize the potentials using the states themselves as an explicit derivation of these potentials, given the states, is straightforward. We fix the energy of the bottom band at $\mathbf{q} = \mathbf{0}$ to zero $(E_0(\mathbf{0}) = 0)$. The corresponding state $\phi_{0, \mathbf{0}}(\mathbf{x})$, being the ground state of the Hamiltonian, has no roots and may in turn be parametrized as 
\begin{equation}
    \phi_{0, \mathbf{0}}(\mathbf{x}) = Ce^{\Lambda(\mathbf{x})}.
\end{equation}
We choose the normalization such that the spatial average $\langle\Lambda(\mathbf{x})\rangle$ vanishes.
\begin{equation}
    \langle\Lambda(\mathbf{x})\rangle\equiv\frac{1}{|D|}\int_D \Lambda(\mathbf{x})\, d\mathbf{x} = 0
\end{equation}
where $D$ denotes a unit cell of the periodic potential. This justifies the name logarithmic fluctuations for the function $\Lambda(\mathbf{x})$. From the Schr\"odinger equation, we may now express the potential as
\begin{equation}\label{potential-Lambda}
    V = \frac12(\nabla^2\Lambda + |\boldsymbol{\nabla}\Lambda|^2)
\end{equation}

Now we turn our attention to building a perturbative formulation of the adjacent energies and states, writing
\begin{equation}\label{stateexpansion}
    \phi_{0, \mathbf{q}}(\mathbf{x}) = \frac{1 + \cdots}{\sqrt{|D|\langle e^{2\Lambda}\rangle}}\exp\left[\Lambda(\mathbf{x}) + iq_i\eta^{(1)}_i(\mathbf{x}) + q_iq_j\eta^{(2)}_{ij}(\mathbf{x}) + iq_iq_jq_k\eta^{(3)}_{ijk}(\mathbf{x}) + q_iq_jq_kq_l\eta^{(4)}_{ijkl}(\mathbf{x}) + \cdots \right]
\end{equation}
We keep the normalization (and overall phase choice) such that $\langle\eta^{(n)}\rangle = 0$ for all $n$. We similarly expand the band structure close to $\mathbf{q} = \mathbf{0}$ as
\begin{equation}\label{energyexpansion}
    E_{0}(\mathbf{q}) = \frac12m^{-1}_{ij}q_iq_j + \frac{1}{24}Q_{ijkl}q_iq_jq_kq_l + \cdots
\end{equation}
This expansion does not need to include odd terms since the Hamiltonian in eq. \eqref{hamiltonian} is time reversal symmetric and therefore $E_\nu(-\mathbf{q}) = E_\nu(+\mathbf{q})$.

Plugging eqs. \eqref{stateexpansion} and \eqref{energyexpansion} into the Schr\"odinger equation
\begin{equation}
    \left[-\frac12(\boldsymbol{\nabla} + i\mathbf{q})^2 + \frac12(\nabla^2\Lambda + |\boldsymbol{\nabla}\Lambda|^2) - E_0(\mathbf{q})\right]\phi_{0, \mathbf{q}} = 0
\end{equation}
we find the perturbative ladder
\begin{equation}\label{eta1}
    \boldsymbol{\nabla}.e^{2\Lambda}\boldsymbol{\nabla}\eta^{(1)}_i = -\partial_ie^{2\Lambda}
\end{equation}
\begin{equation}\label{eta2}
    \boldsymbol{\nabla}.e^{2\Lambda}\boldsymbol{\nabla}\eta^{(2)}_{ij} = e^{2\Lambda}\left(2\partial_i\eta^{(1)}_j + \boldsymbol{\nabla}\eta^{(1)}_i.\boldsymbol{\nabla}\eta^{(1)}_j + \delta_{ij} - m^{-1}_{ij}\right)
\end{equation}
\begin{equation}\label{eta3}
    \boldsymbol{\nabla}.e^{2\Lambda}\boldsymbol{\nabla}\eta^{(3)}_{ijk} = -2e^{2\Lambda}\left(\partial_i\eta^{(2)}_{jk} + \boldsymbol{\nabla}\eta_i^{(1)}.\boldsymbol{\nabla}\eta^{(2)}_{jk}\right)
\end{equation}
\begin{equation}\label{eta4}
    \boldsymbol{\nabla}.e^{2\Lambda}\boldsymbol{\nabla}\eta^{(4)}_{ijkl} = e^{2\Lambda}\left(2\boldsymbol{\nabla}\eta^{(1)}_i.\boldsymbol{\nabla}\eta^{(3)}_{jkl} + 2\partial_i\eta^{(3)}_{jkl} - \boldsymbol{\nabla}\eta^{(2)}_{ij}.\boldsymbol{\nabla}\eta^{(2)}_{kl} - \frac{1}{12}Q_{ijkl}\right)
\end{equation}
Taking the spatial average of eqs. \eqref{eta2} and \eqref{eta4}, we can find formulae for the mass and the quartic dispersion tensor in terms of $\eta^{(1)}$ and $\eta^{(2)}$.
\begin{equation}\label{lambdamass}
    m^{-1}_{ij} = \delta_{ij} + \frac{\langle e^{2\Lambda}\partial_i\eta^{(1)}_j\rangle}{\langle e^{2\Lambda}\rangle}
\end{equation}
\begin{equation}\label{smquartic}
    Q_{ijkl} = -12\frac{\langle e^{2\Lambda}\boldsymbol{\nabla}\eta^{(2)}_{ij}.\boldsymbol{\nabla}\eta^{(2)}_{kl}\rangle}{\langle e^{2\Lambda}\rangle}
\end{equation}
Notably, for one-dimensional systems, an integration by parts can further simplify the formula for mass leading to the simple expression
\begin{equation}\label{1dmass}
    m_\mathrm{1D} = \langle e^{2\Lambda}\rangle\langle e^{-2\Lambda}\rangle
\end{equation}

The quantum metric of the $0$th band at $\mathbf{q} = \mathbf{0}$ may also be written in terms of $\eta^{(1)}$
\begin{equation}\label{smmetric}
    ds^2 = g_{0, ij}(\mathbf{0})dq_idq_j = dq_idq_j\bra{\partial_{q_i}\phi_{0}}(1 - \ket{\phi_0}\bra{\phi_0})\ket{\partial_{q_j}\phi_0} = dq_idq_j\left[\frac{\langle \eta^{(1)}_i\eta^{(1)}_j e^{2\Lambda}\rangle}{\langle e^{2\Lambda}\rangle} - \frac{\langle \eta^{(1)}_i e^{2\Lambda}\rangle\langle\eta^{(1)}_j e^{2\Lambda}\rangle}{\langle e^{2\Lambda}\rangle^2}\right]
\end{equation}

We seek soliton solutions of
\begin{equation}\label{SMGPE}
    \hat{H}\psi - E\psi - |\psi|^{2\sigma}\psi = 0,
\end{equation}
with energy $E = -h^2/2$, in the limit $h\to0$, where the energy approaches the band edge $E_0(\mathbf{0})=0$. We expand the soliton in the Bloch basis as
\begin{equation}\label{smexpansion}
    \psi(\mathbf{x}; -h^2/2) = h^{\frac{1}{\sigma} - d}\sum_\nu\int_B\frac{d\mathbf{q}}{\sqrt{|B|}}\, e^{i\mathbf{q}\cdot\mathbf{x}}\phi_{\nu,\mathbf{q}}(\mathbf{x})\tilde F_\nu(\mathbf{q}/h,h).
\end{equation}
Where $B$ is the first Brillouin zone and
\begin{equation}
    \tilde{F}_\nu(\mathbf{k}, h) = (2\pi)^{-d/2}\int e^{-i\mathbf{k}\cdot\mathbf{X}} F_\nu(\mathbf{X}, h)\, d\mathbf{X}
\end{equation}
are the Fourier transforms of some envelope functions $F_\nu(\mathbf{X}, h)$. Once these envelope functions are determined, the soliton power is given by
\begin{equation}\label{smpower}
    \mathcal{P}\equiv\|\psi\|_2^2 = h^{\frac{2}{\sigma} - d}\sum_\nu\int_{\mathbb{R}^d}d\mathbf{k}\,|\tilde F_\nu(\mathbf{k},h)|^2 + \mathcal{O}(h^\infty).
\end{equation}
More precisely, the exact expression is obtained by restricting the integration domain to $B/h$. Extending the domain to all of $\mathbb{R}^d$ introduces only a superpolynomially small error as $h\to0$. This is because, as we will see, the real-space envelope functions $F_\nu(\mathbf{X}, h)$ are smooth solutions of differential equations, implying that their Fourier transforms $\tilde F_\nu(\mathbf{k},h)$ decay superpolynomially for large $|\mathbf{k}|$.

Substituting eq. \eqref{smexpansion} into eq. \eqref{SMGPE}, it becomes
\begin{equation}
    \begin{gathered}
        \left[1 + \frac{2}{h^2}E_\nu(h\mathbf{k}) - \frac{2}{h^2}E_{0}(\mathbf{0})\right]\tilde{F}_\nu(\mathbf{k}; h) = \frac{2}{|B|^\sigma}\sum_{\nu_0, \cdots, \nu_\sigma, \nu^\prime_1, \cdots, \nu^\prime_\sigma}\int d\mathbf{k}_1\cdots d\mathbf{k}_\sigma d\mathbf{k}_1^\prime\cdots d\mathbf{k}_\sigma^\prime\;\tilde{F}_{\nu_0}(\mathbf{k}_0)\cdots \tilde{F}_{\nu_\sigma}(\mathbf{k}_\sigma)\tilde{F}^*_{\nu^\prime_1}(\mathbf{k}_1^\prime)\cdots \tilde{F}^*_{\nu^\prime_\sigma}(\mathbf{k}^\prime_\sigma)\\
        \int_D\phi_{\nu_0, h[\mathbf{k} + (\mathbf{k}^\prime_1 - \mathbf{k}_1) + \cdots + (\mathbf{k}_\sigma^\prime - \mathbf{k}_\sigma)]}(\mathbf{x})\,\left[\prod_{i=1}^\sigma\phi_{\nu_i, h\mathbf{k}_i}(\mathbf{x})\right]\, \phi^*_{\nu, h\mathbf{k}}(\mathbf{x})\, \left[\prod_{i = 1}^\sigma\phi_{\nu_i^\prime, h\mathbf{k}_i^\prime}^*(\mathbf{x})\right]\,d\mathbf{x}
    \end{gathered}
\end{equation}
The left hand side of this equation is a linear operator that explicitly depends on the energy dispersion of the periodic system while on the right hand side, the nonlinear expression depends on the mixing integrals involving $2\sigma + 2$ Bloch states over the unit cell, reflecting the significance of the Hilbert space geometry of these states.

We attempt to solve these equations order by order in $h$ as $F_\nu(\mathbf{X}, h) = \sum_{n \geq 0} h^nF_{\nu, n}(\mathbf{X})$. The zeroth order envelope is a rescaled version of the unique positive, radially symmetric, and decaying soliton satisfying $\nabla^2 f_\sigma - f_\sigma + f_\sigma^{2\sigma+1} = 0$:
\begin{equation}
    F_{\nu, 0}(\mathbf{X}) = \delta_{\nu, 0}\frac{\sqrt{|B|}} {(2I_0)^{1/2\sigma}(2\pi)^{d/2}} f_\sigma(\sqrt{m}\,\mathbf{X}).
\end{equation}
with $I_0 = \int_D|\phi_{0, \mathbf{0}}|^{2\sigma + 2}d\mathbf{x}$, and $\sqrt{m}$ being the matrix root of the mass tensor $m_{ij}$ from eq. \eqref{lambdamass}.

To find the higher order functions, it is convenient to write them with the same prefactor and scaling as
\begin{equation}
    F_0(\mathbf{X}, h) = \frac{\sqrt{|B|}} {(2I_0)^{1/2\sigma}(2\pi)^{d/2}} \left[f_\sigma(\sqrt{m}\,\mathbf{X}) + hu(\sqrt{m}\mathbf{X}) + h^2v(\sqrt{m}\mathbf{X}) + \cdots\right]
\end{equation}

For the first order correction, the differential equation to solve is
\begin{equation}
    \begin{gathered}
        \left[\Delta - 1 + (2\sigma + 1)f_\sigma^{2\sigma}(\mathbf{X})\right]\Re[u(\mathbf{X})] = 0\\
        \left[\Delta - 1 + f_\sigma^{2\sigma}(\mathbf{X})\right]\Im[u(\mathbf{X})] = 0
    \end{gathered}
\end{equation}
this sets the real part to $\Re[u(\mathbf{X})] = \alpha^i\partial_if_\sigma(\mathbf{X})$ and the imaginary part to $\Im[u(\mathbf{X})] = \beta f_\sigma(\mathbf{x})$ for arbitrary real constants $\alpha^i$ and $\beta$. Nonzero $\alpha^i$ represent a shift of the soliton, while a nonzero $\beta$ represents an overall phase for the soliton. It costs us no generality to set all of these constants to zero and find $u(\mathbf{X}) = 0$. The same gauge freedom exists at all orders of perturbation as the overall phase and the center location may be fixed as functions of $h$.

At the second order, there is a non-trivial right hand side
\begin{equation}\label{smh2}
    \begin{gathered}
        \left[\Delta - 1 + (2\sigma + 1)f_\sigma^{2\sigma}(\mathbf{X})\right] \Re[v(\mathbf{X})] = \frac{1}{12} Q_{ijkl} (\sqrt{m})_{ia}(\sqrt{m})_{jb}(\sqrt{m})_{kc}(\sqrt{m})_{ld}\partial_a\partial_b\partial_c\partial_df_\sigma(\mathbf{X})\\-
        2(2\sigma + 1)f_\sigma^{2\sigma - 1}\left[\langle\eta^{(2)}_{ij}e^{2\Lambda}\rangle - \frac{\langle\eta^{(2)}_{ij}e^{2(\sigma + 1)\Lambda}\rangle}{\langle e^{2(\sigma + 1)\Lambda}\rangle}\right](\sqrt{m})_{ia}(\sqrt{m})_{jb}\left[f_\sigma\partial_a\partial_bf_\sigma + \sigma(\partial_af_\sigma)(\partial_bf_\sigma)\right]
    \end{gathered}
\end{equation}
where $Q_{ijkl}$ is the quartic dispersion tensor found in eq. \eqref{smquartic}. Given $v(\mathbf{X})$, we can write the power of the soliton as
\begin{equation}
    \mathcal{P} = \frac{|B|h^{\frac{2}{\sigma} - d}}{(2I_0)^{1/\sigma}(2\pi)^d\det(\sqrt{m})}\left[\int d\mathbf{X}f_\sigma^2(\mathbf{X})\right]\left[1 + 2h^2\frac{\int d\mathbf{X}\, f_\sigma(\mathbf{X})\Re v(\mathbf{X})}{\int d\mathbf{X}\, f_\sigma^2(\mathbf{X})} + o(h^2)\right]
\end{equation}
Leading to
\begin{equation}\label{smgeneralslope}
    \frac{d\mathcal{P}}{\mathcal{P}dE} = \frac{1/\sigma - d/2}{E} \, - \, 4\frac{\int d\mathbf{X}\, f_\sigma(\mathbf{X})\Re v(\mathbf{X})}{\int d\mathbf{X}\, f_\sigma^2(\mathbf{X})} + \mathcal{O}(h^2)
\end{equation}

For shallow potentials, we can find explicit leading order formulae for the $\eta^{(1)}$ and $\eta^{(2)}$ and in turn for the metric, the mass tensor, and the quartic dispersion tensor in terms of $\Lambda$. They are
\begin{equation}
    \eta^{(1)}_i = -2\Delta^{-1}\partial_i\Lambda + \mathcal{O}(\Lambda^2)
\end{equation}
\begin{equation}
    \eta^{(2)}_{ij} = -4\Delta^{-2}\partial_i\partial_j\Lambda + \mathcal{O}(\Lambda^2)
\end{equation}
\begin{equation}
    m^{-1}_{ij} = \delta_{ij} - 4\langle \Lambda\Delta^{-1}\partial_i\partial_j\Lambda\rangle + \mathcal{O}(\Lambda^3)
\end{equation}
\begin{equation}\label{smweakquartic}
    Q_{ijkl} = 192\langle\Lambda\Delta^{-3}\partial_i\partial_j\partial_k\partial_l\Lambda\rangle + \mathcal{O}(\Lambda^3)
\end{equation}
\begin{equation}
    g_{0, ij}(\mathbf{0}) = -4\langle\Lambda\Delta^{-2}\partial_i\partial_j\Lambda\rangle + \mathcal{O}(\Lambda^3)
\end{equation}
Taking the trace of eq. \eqref{smweakquartic} yields 
\begin{equation}
    \frac{-1}{48}\sum_k \pdv{^4E_0}{q_i\partial q_j \partial q_k \partial q_k}= \frac{-1}{48}\delta_{kl}Q_{ijkl} = -4\langle\Lambda\Delta^{-2}\partial_i\partial_j\Lambda\rangle + \mathcal{O}(\Lambda^3) = g_{0, ij}(\mathbf{0}) + \mathcal{O}(\Lambda^3)
\end{equation}
which is the weak potential result reported in eq. \eqref{metric-quartic}. Assuming an isotropic form for eq. \eqref{smweakquartic} allows for parametrization of both the metric and the quartic tensor in terms of a single scalar $g_{0, \mathbf{0}}$ as
\begin{equation}
    g_{0, ij}(\mathbf{0}) = g_{0, \mathbf{0}}\delta_{ij} + \mathcal{O}(\Lambda^3)
\end{equation}
\begin{equation}
    Q_{ijkl} = \frac{-48}{d + 2}g_{0, \mathbf{0}}(\delta_{ij}\delta_{kl} + \delta_{ik}\delta_{jl} + \delta_{il}\delta_{jk}) + \mathcal{O}(\Lambda^3)
\end{equation}

Similarly, in this regime the other term on the right hand side of eq. \eqref{smh2} admits the representation
\begin{equation}
    \left[\langle\eta^{(2)}_{ij}e^{2\Lambda}\rangle - \frac{\langle\eta^{(2)}_{ij}e^{2(\sigma + 1)\Lambda}\rangle}{\langle e^{2(\sigma + 1)\Lambda}\rangle}\right] = -2\sigma \langle\Lambda\eta^{(2)}_{ij}\rangle + \mathcal{O}(\Lambda^3) = -2\sigma g_{0, \mathbf{0}}\delta_{ij} + \mathcal{O}(\Lambda^3)
\end{equation}

This allows us to simplify the second order envelope correction as
\begin{equation}
    v(\mathbf{X}) = g_{0, \mathbf{0}}\,\alpha(\mathbf{X})
\end{equation}
where $\alpha(\mathbf{X})$ is real and satisfies
\begin{equation}\label{smweakh2}
    \begin{gathered}
        \left[\Delta - 1 + (2\sigma + 1)f_\sigma^{2\sigma}(\mathbf{X})\right] \alpha(\mathbf{X}) = \frac{-12}{d + 2} \Delta^2f_\sigma(\mathbf{X}) + \frac{4\sigma(2\sigma + 1)}{(\sigma + 1)}f_\sigma^\sigma(\mathbf{X})\Delta f_\sigma^{\sigma + 1}(\mathbf{X})
    \end{gathered}
\end{equation}
All of these may be compressed in the result reported in eq. \eqref{gprime}
\begin{equation}\label{smgprime}
    \frac{d\mathcal{P}}{\mathcal{P}dE} = \frac{1/\sigma - d/2}{E} + C(\sigma, d)g_{0, \mathbf{0}} + \mathcal{O}(E, \Lambda^3)
\end{equation}
with
\begin{equation}
    C(\sigma, d) = -4\frac{\int f_\sigma(\mathbf{X})\alpha(\mathbf{X})\, d\mathbf{X}}{\int f_\sigma^2(\mathbf{X})d\mathbf{X}}
\end{equation}

\section{Section 2: Quantum Geometry and the Effective Mass Tensor}

For a Hamiltonian with periodic scalar and vector potentials $H = -\frac12[\boldsymbol{\nabla} - i\mathbf{A}(\mathbf{x})]^2 + V(\mathbf{x})$, the mass tensor and the quantum geometric tensor are both expressible in terms of the vector valued velocity matrix $\boldsymbol{\beta}_{\mu\nu} \equiv \bra{\phi_{\mu,\mathbf{q}}}-i\boldsymbol{\nabla} - \mathbf{A}(\mathbf{x}) + \mathbf{q}\ket{\phi_{\nu,\mathbf{q}}}$ as
\begin{equation}
    (m^{-1}_\mu)_{ij} = \pdv{^2E_\mu}{q_i \partial q_j} = \delta_{ij} - \sum_{\nu\neq \mu} \frac{\beta_{\mu\nu, i}\beta_{\mu\nu,j}^* + \beta_{\mu\nu,i}^*\beta_{\mu\nu,j}}{E_\nu - E_\mu}
\end{equation}
and
\begin{equation}
    (Q_\mu)_{ij} = (g_\mu)_{ij} - \frac{i}{2}(\Omega_\mu)_{ij} = \sum_{\nu\neq \mu}\frac{\beta_{\mu\nu,i}\beta_{\mu\nu,j}^*}{(E_\mu - E_\nu)^2}.
\end{equation}

These imply the matrix inequality
\begin{equation}\label{smbound}
    g_0(\mathbf q) \leq \frac{\mathds{1} - m_0^{-1}(\mathbf{q})}{2\big(E_1(\mathbf q)-E_0(\mathbf q)\big)}.
\end{equation}
which is saturated if and only if all matrix elements $\boldsymbol{\beta}_{\mu0}$ vanish for $\mu > 1$. Equivalently,
\begin{equation}
    [\boldsymbol{\nabla} - i\mathbf{A}(\mathbf{x}) + i\mathbf{q}]\ket{\phi_{0,\mathbf q}} = i\boldsymbol{\beta}_{00}\ket{\phi_{0,\mathbf q}} + i\boldsymbol{\beta}_{10}\ket{\phi_{1,\mathbf q}} .
\end{equation}

Assuming $\mathbf{A}(\mathbf{x}) = \mathbf{0}$, and at the band minimum, $E_0(\mathbf{0}) = 0$, this is equivalent to
\begin{equation}
    [H - E_1(\mathbf{0})]\boldsymbol{\nabla}\ket{\phi_{0,\mathbf 0}} = 0
\end{equation}
Written in terms of the $\Lambda(\mathbf{x}) \equiv \log(\phi_{0, \mathbf{0}}(\mathbf{x})) - \langle\log \phi_{0, \mathbf{0}}(\mathbf{x})\rangle$, this becomes.
\begin{equation}\label{smsaturator}
    E_1(\mathbf{0})\Lambda + \frac12(\Delta\Lambda + |\boldsymbol{\nabla}\Lambda|^2) = \mathrm{const.}
\end{equation}
This is equivalent to eq. \eqref{saturator}. We now focus on the one dimensional case with period $a$. In the absence of the potential, when $V(x) = 0$, the first excited energy is $E^\mathrm{free}_1(0) = 2(\pi/a)^2$. We parametrize the excited energy as $E_1(0) = 2(\pi/a)^2(1 + \kappa^2)$.

In the shallow limit, we write
\begin{equation}
    \Lambda_\kappa(x) = \sum_{n \geq 1} \kappa^n\Lambda_n(x)
\end{equation}
and find at first order
\begin{equation}
    \Lambda_1(x) = \lambda\cos(\frac{2\pi x}{a})
\end{equation}
for some $\lambda$. The second order equation gives
\begin{equation}
    \Lambda_2(x) = -\frac{\lambda^2}{6}\cos(\frac{4\pi x}{a})
\end{equation}
with still no condition on $\lambda$. Finally, the third order equation is
\begin{equation}
    \Lambda_3^{\prime\prime}(x) + (\frac{2\pi}{a})^2\Lambda_3(x) = (\frac{2\pi}{a})^2\lambda\left[(\frac{\lambda^2}{3} - 1)\cos(\frac{2\pi x}{a}) + \frac{\lambda^2}{3}\cos(\frac{6\pi x}{a})\right]
\end{equation}
which admits a periodic solution only if $\lambda = \pm\sqrt{3}$. Apart from a simple shift, both signs describe the same potential and we may write
\begin{equation}
    \Lambda_\kappa(x) = \sqrt{3}\kappa\cos(\frac{2\pi x}{a}) + \mathcal{O}(\kappa^2)
\end{equation}
Using our shallow potential formulae, this yields
\begin{equation}
    m_0(0) = 1 + 6\kappa^2 + \mathcal{O}(\kappa^3)\;\;\;;\;\;\; g_0(0) = 6(\frac{\kappa a}{2\pi})^2 + \mathcal{O}(\kappa^3)
\end{equation}
These indeed saturate the bound \eqref{smbound}.

In the larger $\kappa$ limit, we can have a simple quadratic ansatz $\Lambda = A + Bx^2$ for $-a/2 \leq x \leq a/2$ that solves eq. \eqref{smsaturator} almost everywhere. The zero average condition enforces $A = -Ba^2/12$. To satisfy the equation, the constant $B$ is also set and we find
\begin{equation}\label{deeplambda}
    \Lambda \approx - (\frac{\pi}{a})^2(1 + \kappa^2)(x^2 - \frac{a^2}{12})
\end{equation}
This describes a Gaussian state localized at the center of the unit cell inside a quadratic potential well
\begin{equation}
    V \approx  2 (\frac{\pi}{a})^4(1 + \kappa^2)^2x^2\, - (\pi/a)^2(1 + \kappa^2)
\end{equation}
In this deep lattice regime, the mass grows exponentially (See eq. \eqref{1dmass}). To find the metric, we first simplify eq. \eqref{eta1} to find
\begin{equation}
    \frac{d}{dx}\eta^{(1)}(x) = -1 + \frac{e^{-2\Lambda}}{\langle e^{-2\Lambda}\rangle}
\end{equation}
For localized Bloch states, the second term is negligible where the eigenstate has significant amplitude and therefore $\eta^{(1)}(x) \approx C - x$. This reduces the metric from eq. \eqref{smmetric} to the variance $\sigma^2_x$ of the Bloch state within a unit cell. Comparing with harmonic oscillator results, we find
\begin{equation}
    g_0(0)\approx \frac{(a/2\pi)^2}{1 + \kappa^2}
\end{equation}

To saturate \eqref{smbound} with nonzero magnetic fields, we may still assume the band minimum to occur at $\mathbf{q} = \mathbf{0}$ without loss of generality. Once again, we set the ground state energy to zero $E_0(\mathbf{0}) = 0$. We formulate the Bloch state as 
\begin{equation}
    \phi_{0, \mathbf{0}}(\mathbf{x}) = C \exp[\Lambda(\mathbf{x}) + i\Gamma(\mathbf{x})]
\end{equation}
Where both $\Lambda$ and $\Gamma$ are real valued, periodic, and have zero spatial average. Then, defining the gauge invariant vector field
\begin{equation}
    \mathbf{U}(\mathbf{x}) \equiv \mathbf{A}(\mathbf{x}) - i\boldsymbol{\nabla}\Gamma(\mathbf{x})
\end{equation}
The Schr\"odinger equation sets the scalar potential in terms of $\Lambda(\mathbf{x})$ and $\mathbf{U}(\mathbf{x})$
\begin{equation}
    V(\mathbf{x}) = \frac12\left[\Delta \Lambda(\mathbf{x}) + |\boldsymbol{\nabla}\Lambda(\mathbf{x})|^2 - |\mathbf{U}(\mathbf{x})|^2\right]
\end{equation}
while also implying the constraint
\begin{equation}
    \div e^{2\Lambda(\mathbf{x})}\mathbf{U}(\mathbf{x}) = 0
\end{equation}
Then, the optimality condition
\begin{equation}
    \left[H - E_1(\mathbf{0})\right](\boldsymbol{\nabla} - i\mathbf{A})\phi_{0, \mathbf{0}}(\mathbf{x}) = 0
\end{equation}
turns into the nonlinear pair of conditions
\begin{equation}
    \boldsymbol{\nabla}\left\{\left[E_1(\mathbf{0}) + \frac{\Delta}{2}\right]\Lambda + \frac12|\boldsymbol{\nabla}\Lambda|^2\right\} = \mathbf{U}.\boldsymbol{\nabla}\mathbf{U}
\end{equation}
\begin{equation}
    \left[E_1(\mathbf{0}) + \frac{\Delta}{2}\right] \mathbf{U} = - (\boldsymbol{\nabla}\Lambda . \boldsymbol{\nabla})\mathbf{U} - (\mathbf{U}.\boldsymbol{\nabla})\boldsymbol{\nabla}\Lambda
\end{equation}

In two dimensions, we can find the solution
\begin{equation}
    \Lambda(\mathbf{x}) = 0\;\;\;;\;\;\;\mathbf{U}(\mathbf{x}) = A_0\cos(\frac{2\pi x}{a}) \hat{\mathbf{y}}
\end{equation}
leading to the Hamiltonian
\begin{equation}
    H = -\frac12\nabla^2 + iA_0\cos(\frac{2\pi x}{a}) \partial_y
\end{equation}
The ground state is $\phi_{0, \mathbf{0}}(\mathbf{x}) = C$. The first excited state at $\mathbf{q} = \mathbf{0}$ is $\phi_{1, \mathbf{0}} = Ce^{2i\pi x/a}$ with energy $E_1(\mathbf{0}) = \frac12(2\pi/a)^2$. The metric and the mass tensors for this system are
\begin{equation}
    m_0(\mathbf{0}) = \begin{pmatrix}1 & 0 \\ 0 & \frac{1}{1 - 2(A_0a/2\pi)^2}\end{pmatrix}
\end{equation}
and
\begin{equation}
    g_0(\mathbf{0}) = 2A_0^2(\frac{a}{2\pi})^4\begin{pmatrix}0 & 0 \\ 0 & 1\end{pmatrix}
\end{equation}
which indeed saturate eq. \eqref{smbound}.

\end{document}